\documentclass[12pt]{article}
\usepackage{graphicx}

\usepackage{epsfig,amsmath,amssymb,verbatim,mathrsfs,subfigure}
\def\beq{\begin{eqnarray}}
\def\eeq{\end{eqnarray}}
\def\bea{\begin{eqnarray}}
\def\eea{\end{eqnarray}}

\def\gev{\, {\rm GeV}}

\newcommand{\gsim}{\lower.7ex\hbox{$\;\stackrel{\textstyle>}{\sim}\;$}}
\newcommand{\lsim}{\lower.7ex\hbox{$\;\stackrel{\textstyle<}{\sim}\;$}}

\def\stilde{\widetilde}

\newcommand{\newc}{\newcommand}
\newc{\Nc}{N_{c}}
\newc{\CG}{C_G}
\newc{\gp}{g'}
\newc{\stopi}{\stilde t_i}
\newc{\sboti}{\stilde b_i}
\newc{\staui}{\stilde \tau_i}
\newc{\stopj}{\stilde t_j}
\newc{\sbotj}{\stilde b_j}
\newc{\stauj}{\stilde \tau_j}
\newc{\stopI}{\stilde t_1}
\newc{\stopII}{\stilde t_2}
\newc{\sbotI}{\stilde b_1}
\newc{\sbotII}{\stilde b_2}
\newc{\stauI}{\stilde \tau_1}
\newc{\stauII}{\stilde \tau_2}
\newc{\sstop}{s_{t}}
\newc{\cstop}{c_{t}}
\newc{\ssbot}{s_{b}}
\newc{\csbot}{c_{b}}
\newc{\sstau}{s_{\tau}}
\newc{\cstau}{c_{\tau}}
\newc{\Sstop}{s_{2t}}
\newc{\Cstop}{c_{2t}}
\newc{\Ssbot}{s_{2b}}
\newc{\Csbot}{c_{2b}}
\newc{\Sstau}{s_{2\tau}}
\newc{\Cstau}{c_{2\tau}}
\newc{\salpha}{s_\alpha}
\newc{\calpha}{c_\alpha}
\newc{\Calpha}{c_{2\alpha}}
\newc{\Salpha}{s_{2\alpha}}
\newc{\sbetapm}{s_{\beta_\pm}}
\newc{\cbetapm}{c_{\beta_\pm}}
\newc{\Sbetapm}{s_{2 \beta_\pm}}
\newc{\Cbetapm}{c_{2 \beta_\pm}}
\newc{\sbetaO}{s_{\beta_0}}
\newc{\cbetaO}{c_{\beta_0}}
\newc{\SbetaO}{s_{2 \beta_0}}
\newc{\CbetaO}{c_{2 \beta_0}}
\newc{\vu}{v_u}
\newc{\vd}{v_d}
\newc{\seL}{\stilde e_L}
\newc{\smuL}{\stilde \mu_L}
\newc{\seR}{\stilde e_R}
\newc{\smuR}{\stilde \mu_R}
\newc{\suL}{\stilde u_L}
\newc{\sdL}{\stilde d_L}
\newc{\suR}{\stilde u_R}
\newc{\sdR}{\stilde d_R}
\newc{\scL}{\stilde c_L}
\newc{\ssL}{\stilde s_L}
\newc{\scR}{\stilde c_R}
\newc{\ssR}{\stilde s_R}
\newc{\snue}{\stilde \nu_e}
\newc{\snumu}{\stilde \nu_\mu}
\newc{\snutau}{\stilde \nu_\tau}
\newc{\Gpm}{G^\pm}
\newc{\Hpm}{H^\pm}
\newc{\FFbS}{\overline{FF}S}
\newc{\FFbV}{\overline{FF}V}
\newc{\FSS}{F_{SS}}
\newc{\FSSS}{F_{SSS}}
\newc{\FFFS}{F_{FFS}}
\newc{\FFFbS}{F_{\overline{FF}S}}
\newc{\FSSV}{F_{SSV}}
\newc{\FVS}{F_{VS}}
\newc{\FVVS}{F_{VVS}}
\newc{\FFFV}{F_{FFV}}
\newc{\FFFbV}{F_{\overline{FF}V}}
\newc{\Fgauge}{F_{\rm gauge}}
\newc{\DRbarprime}{$\overline{\rm DR}'$ }
\newc{\DRbar}{$\overline{\rm DR}$ }
\newc{\MSbar}{$\overline{\rm MS}$ }
\newc{\Yu}{{\bf Y}_u}
\newc{\Yd}{{\bf Y}_d}
\newc{\Ye}{{\bf Y}_e}
\newc{\Au}{{\bf a}_u}
\newc{\Ad}{{\bf a}_d}
\newc{\Ae}{{\bf a}_e}
\newc{\bm}{{\bf m}}
\newc{\zhol}{Z^{\rm hol}}

\newc{\rwino}{r_{\tilde W}}
\newc{\rmu}{r_{\tilde H}}
\newc{\ra}{r_A}
\newc{\ccdot}{\!\cdot\!}

\newcommand{\nnmb}{\nonumber}

\newcommand{\lrf}[2]{\left(\frac{#1}{#2}\right)}

\begin{document}

\setlength{\baselineskip}{0.2in}

%#!latex

%\begin{comment}

%@twocolumnfalse\endcsname
%%
%%
\begin{titlepage}
\noindent
\begin{flushright}
{\small CERN-PH-TH/2008-249} \\
{\small MCTP-08-67}  \\
\end{flushright}
\vspace{-.5cm}

\begin{center}
  \begin{Large}
    \begin{bf}
Vacuum Stability with Tachyonic Boundary \\
Higgs Masses in No-Scale Supersymmetry\vspace{0.2cm}\\
or Gaugino Mediation
     \end{bf}
  \end{Large}
\end{center}
\vspace{0.2cm}
\begin{center}
\begin{large}
Jason~L.~Evans$^{a,b}$,
David~E.~Morrissey$^{c}$,  James~D.~Wells$^{a,b}$ \\
\end{large}
  \vspace{0.3cm}
  \begin{it}

$^a$~Michigan Center for Theoretical Physics (MCTP) \\
University of Michigan, Ann Arbor, MI 48109
\vspace{0.2cm}\\
$^b$~CERN, Theory Division, CH-1211 Geneva 23, Switzerland
\vspace{0.2cm}\\
$^c$~Jefferson Laboratory of Physics,
Harvard University \\
Cambridge, Massachusetts 02138
 \vspace{0.1cm}
\end{it}

\end{center}

\center{\today}

%\maketitle
\begin{abstract}

  No-scale supersymmetry or gaugino mediation augmented
with large negative Higgs soft masses at the input scale
provides a simple solution to the supersymmetric flavor problem
while giving rise to a neutralino LSP.  However, to obtain a neutralino
LSP it is often necessary to have tachyonic input Higgs soft masses
that can give rise to charge-and-color-breaking (CCB)
minima and unbounded-from-below (UFB) directions in the low energy
theory.  We investigate the vacuum structure in these theories
to determine when such problematic features are present.
When the standard electroweak vacuum is only metastable, we compute
its lifetime under vacuum tunneling.  We find that
vacuum metastability leads to severe restrictions on the parameter
space for larger $\tan\beta \sim 30$, while for smaller $\tan\beta\sim 10$,
only minor restrictions are found.  Along the way, we derive an exact
bounce solution for tunneling through an inverted parabolic potential.

\end{abstract}

\vspace{1cm}

\end{titlepage}
%\pacs{PACS numbers: }
%]

%\setcounter{footnote}{1}
\setcounter{page}{2}

\tableofcontents

%\end{comment}

%%%%%%%%%%%%%%%%%%%%%%%%%%%%%%%%%%%%%%%%%%%%%%%%%%%%%%%%%%%%%%%%%%%%%%

\section{Introduction}

  Supersymmetry~(SUSY) is a well-motivated extension of the Standard
Model~(SM) that provides an explanation for the stability of the
hierarchy between the weak scale and the Planck
scale~\cite{Martin:1997ns}. However, experiment excludes SUSY from
being an exact symmetry at low energies.  If supersymmetry is
\emph{softly} broken, it is possible to push the superpartner
masses up enough to make the theory consistent with experiment
while still preserving the electroweak-gravity hierarchy. For this
to work, the masses of the SM superpartners should not be
significantly larger than the electroweak scale, putting them
within reach of upcoming particle collider experiments such as the
LHC.

  A major drawback of this scenario of softly-broken low-energy
supersymmetry is that the supersymmetry breaking operators
generically introduce many new sources of quark and lepton flavor
mixing, leading to flavor-changing neutral currents~(FCNC) in
conflict with experiment~\cite{Gabbiani:1996hi}.  One way to cure
this problem is to take No-Scale boundary conditions in which the
soft scalar masses and couplings all vanish at a boundary input
scale, with the gaugino soft masses being the only significant
source of supersymmetry breaking in the visible sector at this
scale~\cite{noscale}. Such boundary conditions can arise from
gaugino mediation~\cite{gaugino-med}, or conformal running
effects~\cite{Nelson:2000sn,confseq,Dine:2004dv,Cohen:2006qc,Murayama:2007ge}.
Low-scale scalar soft terms, which are necessary to obtain a
viable phenomenology, are generated primarily from the gaugino
masses in the course of renormalization group~(RG) running from
the boundary input scale down to near the electroweak scale.  This
dominant contribution is controlled by gauge interactions,
implying that the regenerated scalar soft terms, and thus the
squark and slepton masses, will be approximately flavor-universal
at the low scale.

  A slight generalization of this pure No-Scale picture, which still
leads to scalar soft terms that are roughly flavor universal at
the low scale (for the first and second generations), is to allow
non-vanishing Higgs soft scalar masses at the boundary scale. In
the context of gaugino mediation this can arise if the Higgs
multiplets propagate in the bulk~\cite{Nelson:2000sn}, while for
conformal running, it can emerge from hidden-sector interactions
related to the origin of the $\mu$
term~\cite{Cohen:2006qc,Murayama:2007ge,Perez:2008ng}.
These \emph{Higgs Exempt No-Scale}~(HENS) models often have an
important phenomenological advantage over pure No-Scale
constructions: in large regions of the parameter space HENS models
have a neutralino lightest superpartner particle~(LSP)~\cite{Evans:2006sj}.
This differs from the situation in strict No-Scale models where typically the
predominantly right-handed stau is the lightest SM superpartner
particle~\cite{Schmaltz:2000gy}. While such models can be viable
if the true LSP is a gravitino~\cite{Buchmuller:2005ma}, a
neutralino LSP with exactly
conserved $R$-parity tends to make for a better thermal dark
matter candidate.

  HENS models can push the right-handed stau mass above that of
the lightest neutralino by the influence of the non-vanishing
Higgs soft masses on the RG running of the other soft parameters.
The primary contribution to the slepton soft terms comes from the
hypercharge $D$ term $S_Y= Tr(Y\,m^2) \simeq
(m_{H_u}^2\!-\!m_{H_d}^2)/2$ that appears in the RG equations for all
the MSSM scalar soft
masses~\cite{Martin:1993zk,Kaplan:2000av,Schmaltz:2000gy}. This
contribution increases the right-handed slepton soft masses in the
course of running down provided $S_Y$ is negative. In the portion
of the HENS model parameter space that is also consistent with
electroweak symmetry breaking, the Higgs scalar soft masses must
typically be very tachyonic at the boundary scale to obtain a
neutralino LSP, and the uncolored sfermions still end up being
relatively light~\cite{Evans:2006sj}.

  While these HENS models are compelling, the tachyonic Higgs soft
masses at the boundary scale lead to concerningly large tachyonic low-scale
Higgs soft masses.  These, combined with the small (but positive) slepton
soft masses, suggest that the true vacuum of the theory might be a
charge-and-color-breaking~(CCB) minimum, or there may exist
unbounded-from-below~(UFB) directions that are only stabilized far out
in field space by higher-dimensional operators~\cite{Gunion:1987qv,
Claudson:1983et,Casas:1995pd,Riotto:1995am,Kusenko:1996jn,Ellis:2008mc}.
The presence of
such features need not exclude these regions of the parameter space
provided our SM electroweak vacuum is metastable and long-lived
relative to the age of the universe.

  In the present work we study the vacuum structure of HENS models,
and compute the lifetime of the SM vacuum state when it turns out
to be metastable.  We concentrate on the stability of the SM
vacuum at zero temperature with respect to vacuum tunneling.
Thermal effects in the early universe can also potentially induce
thermal transitions between different vacua, and modify the shape
of the potential itself.  However, these thermal effects tend to
stabilize the origin of the field space to which the SM vacuum is
connected, favoring this vacuum over others that lie further out
in the field space~\cite{Quiros:1994dr}. Thermal effects are
especially effective in delaying the formation of vacua that break
color due to the large thermal corrections from the strong gauge
and top quark Yukawa
couplings~\cite{Quiros:1994dr,Kusenko:1996jn,Carena:1996wj}.
Scalar field evolution during and after inflation may also
populate non-standard vacua, although the precise result depends
on the details of inflation and lies beyond the scope of this
paper~\cite{Dine:1995uk,Ellis:2008mc}. By focusing solely on the
$T=0$ constraints we obtain conservative and unambiguous bounds on
the HENS parameter space that do not depend on the cosmological
history.

  Motivated by gauge coupling unification,
throughout this work we take the input scale to be
$M_c = 10^{16}\,\gev$ and we assume a universal
gaugino mass $m_{1/2}$ at this scale.  The set of independent free
parameters after imposing consistent electroweak symmetry breaking
is therefore\footnote{The sign of the $\mu$ term is also free,
but we fix it to be positive.  This slightly increases the model prediction
for $(g\!-\!2)_{\mu}$, as favored by the experimentally measured
value of this quantity~\cite{Evans:2006sj}.}
\beq
m_{1/2},~~~\tan\beta,~~~m_{H_u}^2(M_c),~~~m_{H_d}^2(M_c).
\label{inputpars}
\eeq
Investigations of No-Scale models with non-universal gaugino masses
can be found in Refs.~\cite{Komine:2000tj,Balazs:2003mm}.
Similar input soft terms have also been studied
in the context of Refs.~\cite{Ellis:2002wv,Baer:2005bu}

  The outline of this paper is as follows.
In Section~\ref{hensvac} we investigate the vacuum
structure of HENS models.  We discuss the general features
of vacuum tunneling and outline the relevant criteria we
use to judge which of the possible CCB and UFB vacua
are the most dangerous in Section~\ref{bounce}.
In Section~\ref{constraint} we map out the portions
of parameter space that are phenomenologically consistent
and give rise to a SM vacuum state that is sufficiently long-lived.
Section~\ref{conc} is reserved for our conclusions.
In Appendix~\ref{appa} we outline our numerical technique
for estimating the vacuum tunneling rate.  We present in Appendix~\ref{appb}
an exact tunneling solution for an inverted-parabola barrier.

%%%%%%%%%%%%%%%%%%%%%%%%%%%%%%%%%%%%%%%%%%%%%%%%%%%%%%%

\section{UFB Directions and CCB Minima in the HENS Model\label{hensvac}}

  With the HENS model input parameters of Eq.~\eqref{inputpars},
relatively small slepton soft masses as well as large tachyonic
Higgs soft masses obtain near the electroweak scale in much of
the phenomenologically allowed parameter space~\cite{Evans:2006sj}.
This spectrum of soft parameters frequently implies the existence
of UFB directions and CCB minima that are much deeper than the
standard electroweak vacuum. Indeed, we find that nearly the
entire allowed parameter space in the HENS model (before imposing
vacuum stability constraints) has at least one UFB direction.
We investigate the existence and nature of such potentially dangerous
vacuum features in the present section.

  Given the soft breaking spectrum that arises in the HENS scenario,
the results of Ref.~\cite{Casas:1995pd} suggest that the most
dangerous vacuum feature will be a sleptonic UFB-3 direction. This
direction has $H_d=0$, with $\tilde \tau_L$, $\tilde \tau_R$,
$\tilde \nu_{L_{i\neq 3}}$, and $H_u^0$ all non-zero. Turning on
expectation values for these fields, there exists a $D$- and
$F$-flat direction that is only lifted by quadratic supersymmetry
breaking operators.  To obtain $F$-flatness, only the $F$-term of
$H_d$ must be cancelled.  This can be arranged by taking
\beq
|\tilde \tau| = |\tilde \tau_L|=|\tilde\tau_R| =
\sqrt{\left|\frac{\mu}{y_{\tau}}\,H_u^0\right|},
\label{stauufb}
\eeq
with the relative phase of $\tilde \tau_L$ and
$\tilde \tau_R$ chosen appropriately.  $D$-flatness is then
obtained by setting
\beq
|\tilde \nu_{L_{i\neq 3}}|^2 =
-\lrf{4\,m_{L_i}^2}{g^2+g'^2} + |\tilde \tau|^2 + |H_u^0|^2,
\eeq
and represents the lowest-energy $F$-flat field configuration provided
\beq
|H_u^0| > \sqrt{\left|\frac{\mu}{2y_{\tau}}\right|^2
+\frac{4m_{L_i}^2}{(g^2+g'^2)}}-\left|\frac{\mu}{2y_{\tau}}\right|.
\label{hulow}
 \eeq
The scalar potential along this direction in field space
then becomes~\cite{Casas:1995pd}
\beq
V_{UFB-3} =
(m_{H_u}^2+m_{L_i}^2)|H_u^0|^2
+\left|\frac{\mu}{y_{\tau}}\right|\,(m_{L_3}^2+m_{E_3}^2+m_{L_{i}}^2)\,|H_u^0|
-\frac{2m_{L_i}^4}{g^2+g'^2}.
\label{vufb3}
\eeq
When $(m_{H_u}^2+m_{L_i}^2)$ is negative, the potential becomes
unbounded in the limit $|H_u^0|\to \infty$.  It will ultimately be
stabilized by loop corrections or higher-dimensional operators
(that we have not included in Eq.~\eqref{vufb3}) at a location
that is very deep and far out in field space relative to the
electroweak vacuum.

  This sleptonic UFB-3 direction is particularly dangerous in the HENS
models on account of the large and negative values of $m_{H_u}^2$
and the smaller values of $m_{L_i}^2$ and $m_{E_i}^2$ that emerge
in the low-energy spectrum.  These properties imply that the
barrier against tunneling from the electroweak vacuum near the
origin out to the deeper UFB-3 direction, arising from the linear
term in $|H_u^0|$ in Eq.~\eqref{vufb3}, will not be especially
large. The barrier will be further weakened by larger values of
$\tan\beta$ which enhance the coupling $y_{\tau} =
m_{\tau}/v\cos\beta$.
Other similar UFB-3 directions may be present in the theory, but
they will generally have larger barriers due to the larger values
of the squark soft masses or the smaller values of the first- and
second-generation lepton Yukawa couplings.

  When $|H_u^0|$ does not satisfy the bound given in Eq.~\eqref{hulow},
the lowest-energy $F$-flat direction in the potential has
$|\tilde{\nu}_{L_i\neq 3}| = 0$, and
is given by~\cite{Casas:1995pd}
\begin{equation}
V_{UFB-3} = m_{H_u}^2|H_u^0|^2
+\left|\frac{\mu}{y_{\tau}}\right|(m_{\tilde \tau_L}^2+m_{\tilde
\tau_R}^2)|H_u^0|
+ \frac{1}{8}(g_1^2+g_2^2)\,\left(|H_u^0|^2
+ \left|\frac{\mu}{y_{\tau}}\right||H_u^0|\right)^2.
%+|\mu H^0_u- y_\tau \tilde\tau^2|^2,
\label{vufb3small}
\end{equation}
This potential is no longer $D$-flat, and is stabilized at a finite
value of $|H_u^0|$.  If this point occurs with $|H_u^0|$ less than
the bound of Eq.~\eqref{hulow}, it is a constrained local
CCB minimum.  On the other hand, when this constrained local extremum
has $|H_u^0|$ larger than the bound of Eq.~\eqref{hulow}, it represents
a saddle point that is unstable under flowing to a non-zero value of
$|\tilde \nu_{L_{i\neq 3}}|$.

  In practice, we find that for smaller values of $|H_u^0|$ the potential
can be reduced further by relaxing the $F$-flatness constraint of
Eq.~\eqref{stauufb}.
%Thus, in our numerical analysis to follow,
%we consider the full potential
%\bea
%V &=& m_{H_u}^2|H_u^0|^2
%+(m_{\tilde \tau_L}^2+m_{\tilde\tau_R}^2)|\tilde\tau|^2
%+ m_{L_{i\neq 3}}^2|\tilde \nu_{L_{i\neq 3}}|^2
%\label{vccb4}
%\\
%&&+ \frac{1}{8}(g_1^2+g_2^2)(|H_u^0|^2 +|\tilde\tau|^2
%-|\tilde \nu_{L_{i\neq 3}}|^2)^2 + |\mu H^0_u- y_\tau \tilde\tau^2|^2.
%\nnmb
%\eea
The effect of dropping the $F$-flatness constraint is that the
minimal potential for a given value of $|H_u^0|$ is deformed
slightly away from the precise UFB-3 form of
Eqs.~(\ref{vufb3},\ref{vufb3small}), but that the same qualitative
features remain.  In particular, there usually remains a local
extremum at $|H_u^0|\neq 0$ and $|\tilde \nu_{L_{i\neq 3}}| = 0$.
We shall designate this local extremum as CCB-4. If this extremum
occurs at smaller values of $|H_u^0|$, on the order of the
electroweak scale, it can be a local CCB minimum. When the CCB-4
extremum occurs with a value of $|H_u^0|$ much larger than the
electroweak scale, it is generally a saddle point that flows in the
$\tilde\nu_{L_{i\neq 3}}$ direction to a genuine UFB-3 direction
of the form given in Eq.~\eqref{vufb3}.  Even when it is only a
saddle, the CCB-4 point plays an important role in determining the
tunneling rate from the electroweak minimum to the UFB-3
direction, as we will discuss below.

  A stable CCB-4 minimum with stops can also arise for smaller values
of $\tan\beta$ and a correspondingly larger $y_t$ Yukawa coupling.
In this case we have $H_u=0$ while $|\tilde t_L|=|\tilde t_R|
=|\tilde t|$ and $H_d^0$ are all non-zero.  The relevant potential is
\begin{equation}
V_D=m_{H_d}^2|H_d^0|^2 + (m_{\tilde{t}_L}^2+m_{\tilde{t}_R}^2)|\tilde t|^2
+ |\mu\,H_d-y_t\tilde t^2|^2
+\frac{g^2+g'^2}{8}\left(|H_d|^2+|\tilde t|^2\right)^2.
\end{equation}
This potential is generally stable against excursions in the
$|\tilde d| = |\tilde{d}_{L_{i\neq 3}}| = |\tilde{d}_{R_{i\neq 3}}|$
direction on account of the larger squark soft masses that arise in
the HENS model.  With $m_{H_d}^2 < 0$ at the low scale, this potential
often has a CCB-4 minimum with both $|H_d^0|$ and $|\tilde t|$ non-zero.
However, as we show below, the barrier to tunneling to this
minimum from the standard electroweak minimum to this CCB-4 minimum
is usually safely large, again on account of the larger values of
the squark soft masses as well as the less negative values of $m_{H_d}^2$.
For similar reasons, we expect that the rate for tunneling to the other
potential CCB minima discussed in Ref.~\cite{Casas:1995pd}
will typically be less constraining than the rate to tunnel to
a stau UFB-3 direction.

%%%%%%%%%%%%%%%%%%%%%%%%%%%%%%%%%%%%%

\section{Computing the Vacuum tunneling Rate\label{bounce}}

  The existence of vacua deeper than the standard electroweak minimum
in HENS models implies there is a danger of tunneling into one of these
phenomenologically unacceptable states. At the very least, the lifetime
for this tunneling must be greater than the age of the universe.
The vacuum-to-vacuum transition rate associated with tunneling can be
calculated using path integral methods~\cite{Coleman:1977py,Callan:1977pt}.
In the semiclassical approximation, the lifetime of
the vacuum is found to be
\begin{equation}
\frac{1}{\tau V}=\Gamma/V=Ae^{-S_b[\bar{\phi}]}
\label{liftim}
\end{equation}
where $A$ is a dimension-four prefactor to be discussed below,
$\bar{\phi}_i$ denotes the \emph{bounce solution}
for the $i$-$th$ field, and $S_b$ is the Euclidean action,
\begin{equation}
S_b[\bar{\phi}]=\int d^4x_E\left(\left|\nabla\bar{\phi}_i\right|^2
+U(\bar{\phi}_1,...,\bar{\phi}_i)\right)=T[\bar{\phi}_i]+V[\bar{\phi}_i].
\label{BouAct}
\end{equation}
The \emph{bounce solution} for the fields $\bar{\phi}_i$
is the extremum of the Euclidean action that is ${O}(4)$-symmetric
and obeys the the following boundary conditions:
\begin{eqnarray}
\frac{d\bar{\phi}_i(0)}{d\rho}&=&0 \label{bcond1}\\
\lim_{\rho\to\infty}\,\bar{\phi}_i(\rho)&=&\phi_i^{SM}\label{bcond2}
\end{eqnarray}
where $\rho=(\vec{x}^2+t_E^2)^{1/2}$ is the Euclidean
distance. These boundary conditions correspond to a field that
begins in a metastable vacuum and tunnels through a barrier
separating it from a deeper vacuum state, emerging with
zero kinetic energy.  We focus on ${O}(4)$ symmetric
solutions because these are expected to have the least action,
and therefore dominate the tunneling probability~\cite{Coleman:1977th}.

  Even with the simplification of an $O(4)$ symmetry,
the equation of motion for the bounce cannot in general be solved
analytically.  We compute the bounce solution numerically using
the \emph{improved action method}~\cite{Kusenko:1995jv}. The
details of this method are outlined in Appendix~\ref{appa}. Even
more difficult to compute is the non-exponential pre-factor $A$ in
Eq.~(\ref{liftim})~\cite{Callan:1977pt}.  On general grounds, we
expect it to be on the order of the mass scale setting the size of
the potential barrier.  In low-energy supersymmetry, a good
estimate for this number is $(100\gev)^4$, which we take to be the
case throughout the rest of this article.  The precise value of
this pre-factor is unlikely to affect our qualitative conclusions
as the multiplicative uncertainty in its value is much more
slowly-varying than the exponentiated large values of the action
that lead to acceptable lifetimes. Our choice for the pre-factor
is also conservative, in that choosing a larger number here would
only exclude more points. With this pre-factor, it is found that
the lifetime of the SM vacuum will be greater than the age of the
universe, $1/(\Gamma/V) \gtrsim t_0^4$, provided $S_b[\bar{\phi}]
> 400$.

  The value of the bounce action for tunneling between a pair
of local vacua depends on the relative depth of the minima, the
number of distinct minima, the height of the barriers between
them, and the relative size of the field values within them.
In general, the bounce solution represents a configuration of
fields that simultaneously minimizes these
opposing contributions, with the kinetic term favoring slowly
varying fields, and the potential term preferring to reach the
deepest minimum as quickly as possible.  Some intuition for
this balance can be obtained by examining special cases in
which the bounce solution can be derived analytically.
One such example is a potential consisting of piecewise linear
segments~\cite{Duncan:1992ai}.  Motivated by the UFB-3 potential
in Eq.~\eqref{vufb3}, we present another exact tunneling
solution in Appendix~\ref{appb} for an inverted
parabolic potential.\footnote{ While this potential has the same form as the
UFB-3 potential given in Eq.~\eqref{vufb3}, we cannot apply our
exact solution to tunneling through the UFB-3 barrier since the
relevant kinetic terms for this case are non-canonical.}

  As expected, our analytic tunneling solution shows that the bounce
action increases with the size of the barrier.  This solution also
implies that the depth of the minimum to which one is tunneling to
ceases to matter once it becomes very deep.  Thus, we can safely compute
the rate to tunnel into a UFB-3 direction without knowing where it is
ultimately stabilized provided the corresponding minimum is very deep
relative to the height of the barrier.
On the other hand, the relative depth of the minima is relevant
to the tunneling rate when they are nearly degenerate,
as can be seen from the analytic \emph{thin-wall} approximate
solution that can be safely applied in this case~\cite{Coleman:1977py}.
One important feature not captured by our one-dimensional analytic tunneling
solution is that the bounce action tends to also increase when there
are more independent fields involved in the tunneling process,
as each one of them contributes non-negatively to the kinetic portion
of the bounce action. For these reasons, numerical solutions are used
to get the details correct.

%%%%%%%%%%%%%%%%%%%%%%%%%%%%%%%%%%%

\section{Vacuum Stability Bounds on the HENS Parameters\label{constraint}}

  The possibility of tunneling from the SM vacuum to a phenomenologically
unacceptable vacuum places strong constraints on the parameter space
of the HENS model.  To investigate these constraints, we have calculated
the bounce action for tunneling to a CCB-4 minimum or a UFB-3 direction
for a series of representative HENS parameter sets.  Our strategy is to
fix $m_{1/2}$ and $\tan\beta$, and scan over the input values
of $m_{H_u}^2(M_c)$ and $m_{H_d}^2(M_c)$ that lead to an acceptable
low energy spectrum.  We focus on the
values $m_{1/2}=500,1000\gev$ for $\tan\beta=30$ and
$m_{1/2}=300,500\gev$ for $\tan\beta=10$.  In each of these scans
we have consider points that meet the phenomenological constraints
laid out in Ref.~\cite{Evans:2006sj}, including the current LEP lower
bounds on the
lightest Higgs boson. We also keep the points with a charged
slepton as the lightest MSSM superpartner found in Ref.~\cite{Evans:2006sj},
noting that they would require a more complicated cosmology and likely
R-parity violation or
a gravitino LSP~\cite{Buchmuller:2005ma}.  By keeping such points,
our results are also applicable to minimal gaugino mediation subject
to our assumptions about the gaugino masses and the input
(compactification) scale.

\begin{figure}[ttt]
\begin{center}
  \includegraphics[width=0.7\textwidth]{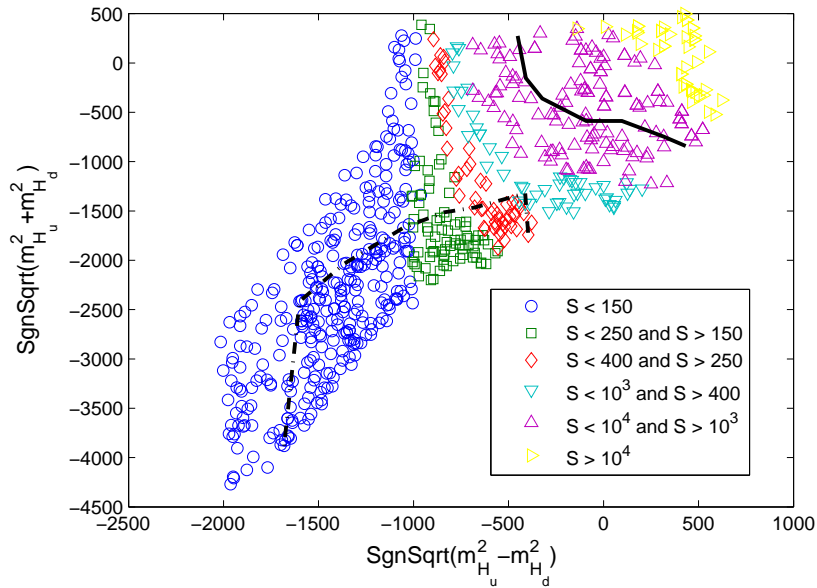}
\vspace{0.3cm}
\caption{ The bounce action for tunneling to a stau UFB-3 direction or
a CCB-4 minimum as a function of the HENS model high scale input
parameters $m_{H_u}^2(M_c)$ and $m_{H_d}^2(M_c)$ for $m_{1/2} = 500\,\gev$
and $\tan\beta = 30$.  All points shown are consistent with collider
phenomenology. The points enclosed below by the dash-dot line have a
neutralino LSP.  The solid line separates the region with a CCB-4
minimum or saddle point (left) from that which only has a UFB-3
(right). $S>400$ is cosmologically safe.
\label{m12500b30}}
\end{center}
\end{figure}

\begin{figure}[ttt]
\begin{center}
  \includegraphics[width=0.7\textwidth]{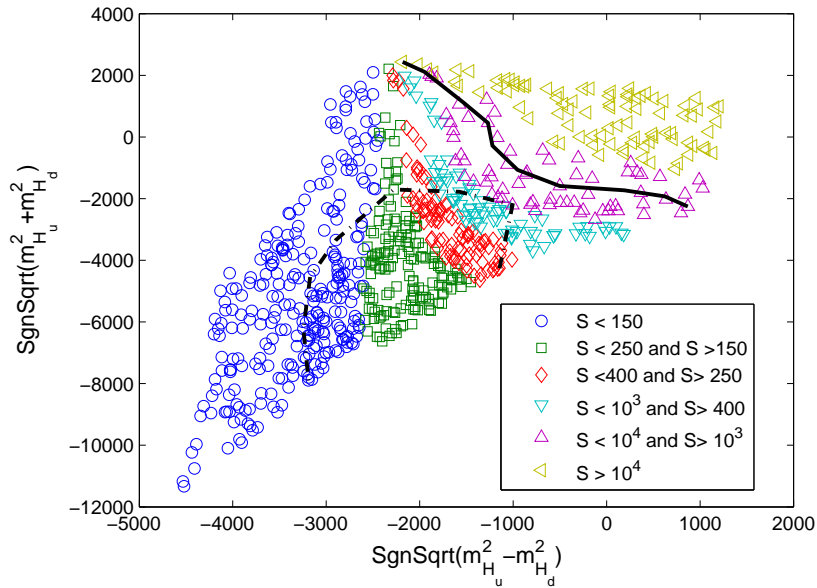}
\vspace{0.3cm} \caption{ The bounce action for tunneling to a
stau UFB-3 direction or a CCB-4 minimum as a function of the
high-scale HENS model input parameters $m_{H_u}^2(M_c)$ and
$m_{H_d}^2(M_c)$ for $m_{1/2}=1000\gev$, $\tan\beta=30$ and
$sgn(\mu)=1$.  All points shown are consistent with collider
phenomenology. The points enclosed below by the dash-dot line have
a neutralino LSP. The solid line separates the region with a CCB-4
minimum or saddle point (left) from that which only has a UFB-3
direction. $S>400$ is cosmologically safe. \label{m121000b30} }
\end{center}
\end{figure}

  We first consider the cases with $\tan\beta = 30$, where the
tau Yukawa coupling is enhanced.  This has the effect of opening
up a CCB-4 minimum or CCB-4-like saddle point that flows to a
UFB-3 direction. In Fig.~\ref{m12500b30} we show ranges of the
bounce action for tunneling out of the SM vacuum to a CCB-4
minimum or a UFB-3 direction for $m_{1/2} = 500\,\gev$ as a
function of the input values of $m_{H_u}^2(M_c)$ and
$m_{H_d}^2(M_c)$. All points to the left of the solid line in this
figure have either a CCB-4 minimum or saddle point. The points to
the right have either a stable SM vacuum or a UFB-3 direction with
no CCB-4 stationary point.  To the left and the right of the solid
line we show the bounce action for tunneling along the direction
leading to the CCB-4 stationary point or to the UFB-3 direction,
whichever is smaller.\footnote{Except for those points very near
the solid line, the CCB-4 extremum is a saddle point flowing to
the UFB-3 direction.} We emphasize the CCB-4 stationary point
here, even when it is only a saddle point, because we find that
the most dangerous lowest-action tunneling path from the
electroweak minimum is typically one that passes near this point
with $\tilde \nu_{L_{i\neq 3}}=0$. The dot-dashed line in
Fig.~\ref{m12500b30} indicates the upper border of the portion of
parameter space in which the LSP is the lightest neutralino. (See
Ref.~\cite{Evans:2006sj} for more details.).  In
Fig.~\ref{m121000b30} we show the same quantities as in
Fig.~\ref{m12500b30} for $m_{1/2} = 1000\,\gev$.

  Both Fig.~\ref{m12500b30} and Fig.~\ref{m121000b30}
show that for $\tan\beta = 30$ the lifetime for tunneling
to a stau UFB-3 direction or CCB-4 minimum is shorter than the
age of the universe, corresponding to $S_b < 400$, over a large portion
of the otherwise acceptable parameter space.  As expected, the newly
disallowed regions are those with large and negative values of
$m_{H_u}^2$ and $m_{H_d}^2$ at the input scale $M_c$. Of these two
soft masses, only $m_{H_u}^2$ appears in the potential relevant
for the UFB-3 direction or the CCB-4 minimum, and it has the stronger effect.
Indeed, the isocontours of $S_b$ coincide roughly with lines
of constant $m_{H_u}^2$ for smaller values of $m_{H_d}^2$.
For smaller values of $m_{H_u}^2$ the SM vacuum becomes sufficiently
long-lived to describe our universe.  The bounce action for these points
is much larger simply because the effective width of the barrier is
larger. Note that this stable region includes the minimal gaugino
mediation point, $m_{H_u}^2= m_{H_d}^2 = 0$ at $M_c$.

  Comparing the plots for $m_{1/2} = 500\,\gev$ and $m_{1/2} =
1000\,\gev$, we see that larger $m_{1/2}$ tends to yield a
slightly more stable SM vacuum. This arises simply because
increasing the input gaugino masses also increases the low-scale
slepton soft masses through RG running.  On the other hand, larger
gaugino masses also permit more negative values of $m_{H_u}^2$, so
there remain significant parameter regions in which the tunneling
rate is too fast.  From Fig.~\ref{m12500b30}, where $m_{1/2} = 500\,\gev$,
we see that nearly the entire region in which the lightest superpartner
is a neutralino is ruled out by our vacuum stability considerations.
For $m_{1/2} = 1000\,\gev$, there is a small region in which the
lightest superpartner is a neutralino and the electroweak vacuum is
sufficiently long-lived.

\begin{figure}[ttt]
\begin{center}
  \includegraphics[width=0.7\textwidth]{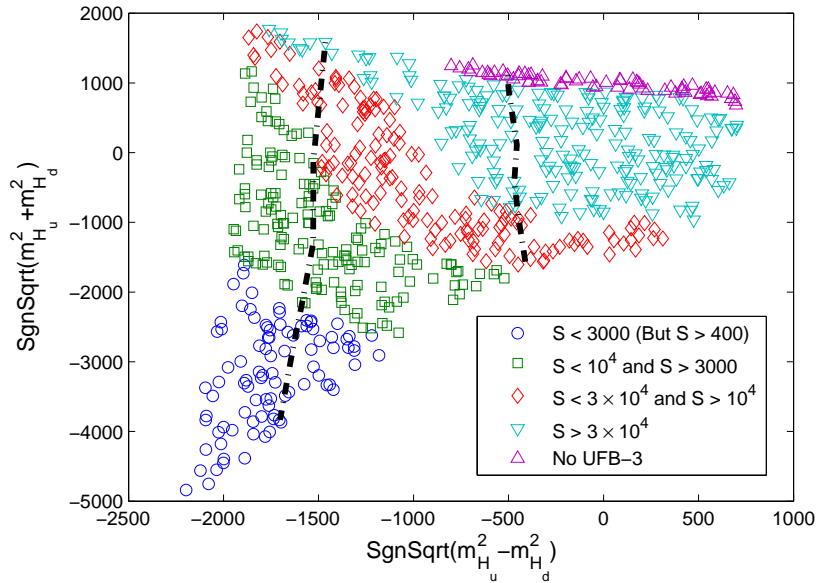}
\vspace{0.3cm} \caption{\label{m12500b10} The bounce action for
tunneling to a stau UFB-3 direction as a function of the HENS model
parameters $m_{H_u}^2$ and $m_{H_d}^2$.  The other HENS parameters
have been fixed to be $m_{1/2}=500\gev$, $\tan\beta=10$, and $sgn(\mu)=1$.
All points shown are consistent with collider phenomenology.
The points between the two dash-dotted lines have a neutralino LSP. $S>400$ is cosmologically safe.}
\end{center}
\end{figure}

\begin{figure}[ttt]
\begin{center}
  \includegraphics[width=0.7\textwidth]{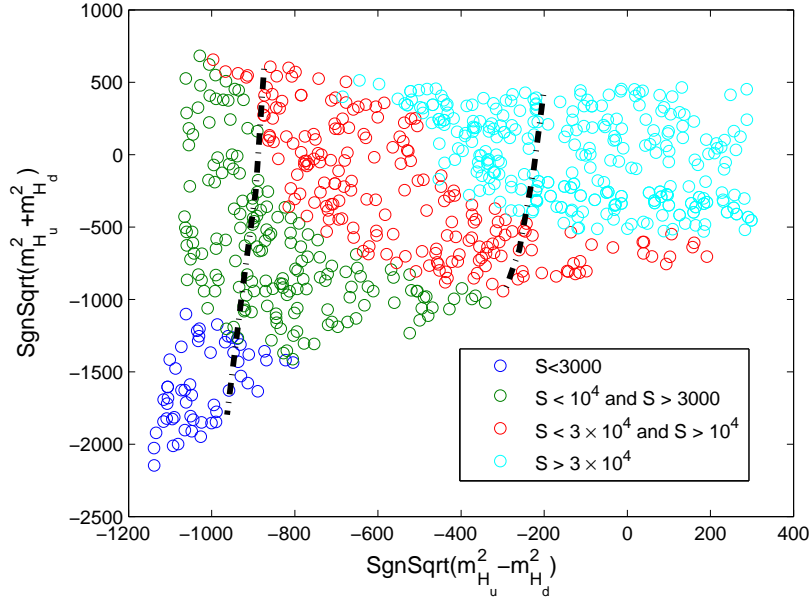}
\vspace{0.3cm} \caption{\label{m12300b10} The bounce action for
tunneling to a stau UFB-3 direction as a function of the HENS model
parameters $m_{H_u}^2$ and $m_{H_d}^2$.  The other HENS parameters
have been fixed to be $m_{1/2}=300\gev$, $\tan\beta=10$, and $sgn(\mu)=1$.
All points shown are consistent with collider phenomenology.
The points between the two dash-dot lines have a neutralino LSP.  $S>400$ is cosmologically safe.}
\end{center}
\end{figure}

 We turn next to the cases with $\tan\beta=10$, for which the
tau Yukawa coupling is less enhanced.  In Figs.~\ref{m12500b10} and
\ref{m12300b10} we show ranges of the bounce
action for tunneling from the SM vacuum to the UFB-3 direction
for $\tan\beta = 10$ and $M_{1/2}=500,\,300\,\gev$ as a function
of the input values of $m_{H_u}^2$ and $m_{H_d}^2$ at $M_c$.
No CCB-4 local extremum is found for any of the points scanned over
since the deviation from $F$-flatness is a sizeable effect in this case.
The dot-dashed lines in these plots enclose the portion of parameter
space in which the LSP is the lightest neutralino.
(See Ref.~\cite{Evans:2006sj} for more details.)
From these plots we see that the lifetime of the standard
electroweak vacuum against tunneling to a UFB-3 direction
is safely large for both $m_{1/2} = 500,\,300\,\gev$.
This is the result of the smaller value of the tau Yukawa
coupling $y_{\tau}$, which gives rise to a larger barrier
against tunneling, Eq.~\eqref{vufb3}. As before, with all else equal $S_b$
increases with
$m_{1/2}$ and so higher values of $m_{1/2}$ remain safe.

\begin{figure}[tth]
\begin{center}
  \includegraphics[width=0.7\textwidth]{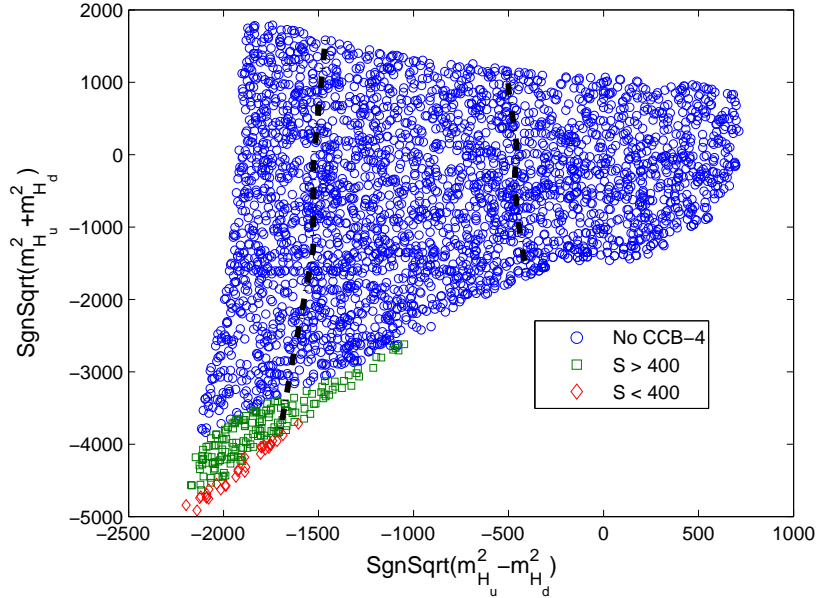}
\vspace{0.3cm} \caption{\label{m12500b10CCB} The bounce action
for tunneling to a stop CCB-4 minimum as a function of the HENS model
parameters $m_{H_u}^2$ and $m_{H_d}^2$.  The other HENS parameters
are $m_{1/2}=500\gev$, $\tan\beta=10$ and $sgn(\mu)=1$.  All these
points are consistent with collider phenomenology.
The points between the two dash-dot lines have a neutralino LSP. $S>400$ is cosmologically safe.}
\end{center}
\end{figure}

  At smaller values of $\tan\beta$ the top quark Yukawa coupling
grows larger, and we should check that the tunneling rate to
CCB-4 minima associated with the stops is adequately small. As
was previously discussed, this direction will only occur if
$m_{H_d}^2$ is large and negative. In Fig.~\ref{m12500b10CCB} we
show contours of the bounce action for tunneling to the stop
CCB-4 minimum, as well as the regions in which the SM minimum is
the true minimum. Only a very few points at the largest and most
negative values of $m_{H_d}^2$ are excluded, while in the great
majority of the parameter space the SM vacuum is the true minimum.
We find a similar result for $m_{1/2} = 300\,\gev$.

\section{Conclusion\label{conc}}

  In the present work we have examined the constraints
placed on the HENS model from vacuum stability.  Due to the large
tachyonic Higgs soft masses that can emerge in this model,
there often arise local vacuum states deeper than the standard electroweak
minimum.  Many points that are consistent with collider
phenomenological constraints (described in Ref.~\cite{Evans:2006sj})
are ruled out because they lead to an overly short-lived SM vacuum.

  The most dangerous vacuum feature is a UFB-3 direction involving
the stau fields.  We have also found a new CCB-4 saddle point that
facilitates tunneling to the UFB-3 direction.
As a result, vacuum tunneling rates tend to be too fast for
larger values of $\tan\beta$.  At lower values of $\tan\beta$,
tunneling to a CCB-4 direction involving stop fields rules out
a very small portion of the parameter space, with the rest of
the parameter space being safely long-lived.

  We conclude that the HENS models with a neutralino LSP and larger values
of $\tan\beta = 30$ are mostly ruled out subject to our assumptions
about the input scale and gaugino mass universality.
On the other hand, minimal gaugino mediation and the HENS models without
a neutralino LSP may still have a sufficiently long-lived electroweak
vacuum state at larger values of $\tan\beta$.  For lower values of $\tan\beta$, such as
$\tan\beta =10$, the constraints from vacuum tunneling are much
weaker, and most of the parameter space that is consistent with collider
lower-energy bounds remains viable.

%%%%%%%%%%%%%%%%%%%%%%%%%%%%%%%%%%%%%%%%%%%%%%%%%%%%%%%%%%%%%%%%%%%%%%
\section*{Acknowledgements}

  We thank F. Adams, P. Kumar, J. Mason, and D. Poland for helpful discussions.
This work is supported in part by the U.S. Department of Energy
under DOE Grant DE-FG02-95ER40899.
J.E. would also like to thank the CERN TH department for its
hospitality while this work was in progress.

%\clearpage
%\newpage

%%%%%%%%%%%%%%%%%%%%%%%%%%%%%%%%%%%%%%%%%%%%%%%%%%%%%%%%%%%%%%%%%%%%%%
%%%%%%%%%%%%%%%%%%%%%%%%%%%%%%%%%%%%%%%%%%%%%%%%%%%%%%%%%%%%%%%%%%%%%%

\appendix

\section{Appendix: The Improved Action Method\label{appa}}

  The bounce action is a stationary point of the Euclidean action
given in Eq.~\eqref{BouAct} subject to the boundary conditions
of Eqs.~(\ref{bcond1},\ref{bcond2}).  The corresponding equations of motion
for the $O(4)$ symmetric solution are
\begin{equation}
%\bigtriangledown^2\bar{\phi}_i=
\frac{d^2\bar{\phi}_i}{d\rho^2}
+ \frac{3}{\rho}\frac{d\bar{\phi}_i}{d\rho}
=
\frac{\partial}{\partial
\bar{\phi}_i}U(\bar{\phi}_i).
\end{equation}
%where $\bigtriangledown^2$ is the four dimensional Laplacian.
where $i$ runs over the independent fields.
These equations are a set of non-linear coupled differential
equations with an {\it{a priori}} unknown starting point.
These conditions together make it a very difficult problem
to solve and require numerical techniques~\cite{Claudson:1983et,
Kusenko:1995jv,Konstandin:2006nd}.

  The technique we use in the present work is called the
\emph{improved action method}~\cite{Kusenko:1995jv}.
In this method, additional terms are added to the action
that are identically zero for the bounce solution.
The advantage of adding these terms is that they make
the bounce solution a minimum of this modified action and not
just an extremum.  The term that does this is found by making the
change of variable $x\to \sigma x$ in Eq. (\ref{BouAct}). Because
the bounce is the extremum of the action, the first derivative of
the scaled action with respect to $\sigma$ will be zero for
$\sigma=1$. This gives the following condition:
\begin{equation}
\left(\sigma^2T[\bar{\phi}_i]+2\sigma^4U[\bar{\phi}_i]\right)|_{\sigma=1}
=0
\label{ScaInv}
\end{equation}
This relation illustrates that the potential term must be
negative. The kinetic term cannot be negative because it is the
integral of a sum of squares. Since the potential term scales as
$\sigma^4$ and the kinetic term scales as $\sigma^2$, Eq.~\eqref{ScaInv}
defines a maximum.  Thus, we have determined the maximal direction of the
saddle point. By adding to the action the absolute value of this
quantity to a positive power, the saddle point of the action
can be turned into a minimum. In this case the improved action is
\begin{equation}
S[\bar{\phi}_i]=T[\bar{\phi}_i]+U[\bar{\phi}_i]+\lambda
\left|T[\bar{\phi}_i]+2U[\bar{\phi}_i]\right|^n\label{ModAct}
\end{equation}
where $\lambda$ and $n$ are positive constants.

  To solve for the bounce with this improved action, we take an
initial profile for the vevs $\phi_i$ with the kinetic and
potential terms
\begin{eqnarray}
T[\phi_i]=2\pi^2\Delta^4\sum_{m=1}^{L-1}
(\rho_{m+1}-\rho_m)\rho_m^3\left[\sum_{i}^n
\frac{(\phi_i^{m+1}-\phi_i^m)^2}{2(\rho_{m+1}-\rho_m)^2\Delta^2}\right]
\\
U[\phi_i]=2\pi^2\Delta^4\sum_{m=1}^{L-1}
(\rho_{m+1}-\rho_m)\rho_m^3U(\phi_{1}^m,...,\phi_{n}^m).
\end{eqnarray}
$\Delta$ is a parameter determined by Eq.~\eqref{ScaInv}.
Inspired by the thin-wall approximation,
we take the following initial guess for the bounce solution
\begin{equation}
\phi(\rho)=a\tanh(b(\rho-\rho_0))+c. \label{thiwall}
\end{equation}
The coefficients $a$ and $c$ can be solved for by applying the
boundary conditions $\phi(0)=\phi_e$ and $\phi(\infty)=\phi_f$,
where $\phi_f$ are the field values in the SM minimum and $\phi_e$
are the values in the vacuum to which the tunneling connects.
Since $\phi_e$ is \emph{a priori} unknown, this leaves $\phi_e$, $b$, and
$\rho_0$ as free parameters. These parameters are determined by
first substituting the field profile in Eq.~\eqref{thiwall} into
the modified action for each $\phi_i$.

  The $\phi_e$ will not in general be the field configuration
of the minima, but rather some field points inside the well. In
the case of a UFB-3 direction there is no minimum, and $\phi_e$
will be some point on the runaway downslope with a potential
energy less than that at SM minimum. The exact value of $\phi_e$
as well as $b$ and $\rho_0$ are determined using a minimization
routine that finds the coefficients that minimize the modified
action. In Ref.~\cite{Kusenko:1996jn} the authors used the
thick-wall approximation as an initial guess. In our case, the
guess given in Eq.~\eqref{ScaInv} works better numerically because
it is adaptable to both thick- and thin-wall potential profiles.
Once the initial profile, coefficients and all, is determined,
we randomly vary each lattice site. The variations are stopped
when further iterations do not reduce the modified action.
To ensure that we arrive at the bounce solution, we choose a
value of $\lambda$ that ensures $0.999<(-T[\phi_i]/2/V[\phi_i])^{1/2}<1.001$.
The smallest value of $\lambda$ able to maintain this condition and used to
optimize the code was close to $0.5$ with $n=1$.

\section{Appendix: Tunneling Through an Inverted Parabola\label{appb}}

  We exhibit here an exact bounce solution for a single field
tunneling through an inverted parabolic potential. To the best of
our knowledge, this simple solution has not been presented
elsewhere in the literature, although a related solution for
a linearized potential was obtained in the pioneering work
of Ref.~\cite{Coleman:1977py}.

  Consider the run-away potential
\beq
V(\phi) = a|\phi| - b|\phi|^2,
\eeq
where $\phi$ is a real scalar field.
The origin is a metastable minimum for $a,\,b>0$.
Focussing on $\phi >0$, the equation of motion
for an $O(4)$-symmetric bounce is
\beq
\frac{d^2\phi}{d\rho^2} + \frac{3}{\rho}\,\frac{d\phi}{d\rho}
+ (2b\,\phi -a) = 0,
\label{del}
\eeq
where $\rho$ is the Euclidean distance.
The boundary conditions for the bounce are
\bea
\frac{d\phi}{d\rho}(\rho=0) &=& 0\label{bc1}\\
\phi(R) &=& 0\label{bc2}\\
\frac{d\phi}{d\rho}(\rho=R) &=& 0\label{bc3}
\eea
For potentials in which the false vacuum is flat, $R\to \infty$,
as we assumed in Eq.~\eqref{bcond2}.
Here, the potential has a singular first derivative at the false
vacuum at the origin.  As a result, the bounce reaches the origin
at finite $R$ and remains there asymptotically~\cite{Duncan:1992ai}.
We have checked that deforming the potential into a concave parabola
very near the origin to resolve this singularity leads to $R\to \infty$,
but otherwise has only a small (and smooth) effect on the solution
presented here away from the origin.

  Eq.~\eqref{del} is a linear second-order differential equation
that can be solved analytically.  The general solution
before imposing boundary conditions is
\beq
\phi(\rho) = \frac{a}{2b} + \frac{1}{x}\left[
c_1\,J_1(x) + c_2\,Y_1(x)\right],
\eeq
where $J_1$ and $Y_1$ are Bessel functions of the first kind,
\beq
x = \frac{\rho}{\sqrt{2b}},
\eeq
and $c_1$ and $c_2$ are constant coefficients.
 The values of $c_1$ and $c_2$, as well as $R$,
are fixed by the boundary conditions, Eqs.~(\ref{bc1}-\ref{bc3}).
The first of these, Eq.~\eqref{bc1}, gives
\beq
0&=& \left.\frac{d}{dx}\left(\frac{1}{x}\left[c_1J_1(x)+c_2Y_1(x)\right]
\right)\right|_{x=0}\nnmb\\
  &=& -\lim_{x\to 0}\left(\frac{1}{x}\left[c_1J_2(x) + c_2\,Y_2(x)\right]
\right),
\eea
where we have made use of the recursion properties of Bessel
functions~\cite{mwbook}.
For this limit to be non-singular, we must have $c_2=0$.  On the
other hand, $J_2(x)/x$ vanishes at the origin so no constraint is
imposed on $c_1$ by this condition.  Applying Eq.~\eqref{bc3},
we find
\beq
0 &=& c_1\left.\frac{d}{dx}\left[\frac{1}{x}J_1(x)\right]\right|_{x=x_+}\\
&=& -c_1\,\frac{1}{x_+}J_2(x_+),\nnmb
\eeq
where we have defined $x_+ = R/\sqrt{2b}$.  From this, we conclude
that $x_+$ must be a zero of $J_2$.  In particular, for the minimal
action bounce solution, it should be the first non-trivial zero:
\beq
x_+ \simeq 5.1356\ldots
\eeq
Finally, let us apply the condition of Eq.~\eqref{bc2} to fix $c_1$,
\beq
c_1 = -\frac{a}{2b}\left[\frac{x_+}{J_1(x_+)}\right].
\eeq

  We can use this exact solution to compute several interesting
quantities related to the bounce.  The bounce action is found to be
%
\begin{comment}
The pieces that go into it are:
\bea
%
\int_0^Rd\rho\,\rho^3 &=& \frac{1}{4b^2}\,\int_0^{x_+}dx\,x^3,\\
%
\frac{1}{2}\left(\frac{d\phi}{d\rho}\right)^2 &=&
c_1^2\,b\,\left[\frac{J_2(x)}{x}\right]^2,\\
%
V(\phi) &=& \frac{a^2}{4b} -c_1^2\,b\,\left[\frac{J_1(x)}{x}\right]^2
\eea
Using Eq.~\eqref{integ}, the bounce action is given by
\end{comment}
%
\bea
S_E &=& 2\pi^2\int_0^Rd\rho\,\rho^3
\left[\frac{1}{2}\left(\frac{d\phi}{d\rho}\right)^2
+V(\phi)\right]\label{se}\\
%
%&=& \frac{2\pi^2}{4b^2}\frac{a^2}{4b}\left(
%\frac{1}{2}x_+^4 -\frac{1}{4}x_+^4\right)\nnmb\\
%
&=& \frac{\pi^2}{32}\,x_+^4\,\frac{a^2}{b^3}\nnmb\\
&\simeq& (215)\,\frac{a^2}{b^3}.\nnmb
\eea
The value of the field $\phi$ at the escape point where it
emerges from tunneling is given by
\beq
\phi(0) = \frac{a}{2b}
-\frac{a}{2b}\left[\frac{x_+}{J_1(x_+)}\right]\,
\lim_{x\to 0}\frac{J_1(x)}{x} \simeq \frac{4a}{b}.
\eeq
For comparison, the maximum of the potential occurs at
$\phi = a/2b$ and the width of the potential barrier is
$\Delta\phi = a/b$.  Thus, the escape point is well beyond the peak,
but only by a factor of a few.

  The exact solution presented here can also be extended to
potentials that are piecewise segments of parabolas, in analogy
to the solution for piecewise-linear potentials presented in
Ref.~\cite{Duncan:1992ai}.  (When the local potential is a concave parabola,
the general local solution consists of modified Bessel functions
of the first kind.)  This offers the possibility of obtaining
closed-form expressions for bounce solutions to a
wide variety of potentials by approximating them with segments
of parabolas.  In practice, we find that the matching conditions
between adjacent parabolic segments often lead to complicated implicit
transcendental equations.  The generalization to thermal transitions
over piecewise-parabolic barriers is also straightforward.

%\clearpage

%%%%%%%%%%%%%%%%%%%%%%%%%%%%%%%%%%%%%%%%%%%%%%%%%%%%%%%%%%%%%%%%%

\end{document}